# On the ferroelectric and magnetoelectric mechanisms in low $Fe^{3+}$ doped $TbMnO_3$

R. Vilarinho,[1] E. Queiros,[2] D. J. Passos,[1] D. A. Mota,[1] P. B. Tavares,[2] M. Mihalik jr.,[3] M. Zentkova,[3] M. Mihalik,[3] A. Almeida,[1] and J. Agostinho Moreira[1]

[1]IFIMUP and IN-Institute of Nanoscience and Nanotechnology, Physics and Astronomy Department of Faculty of Sciences of University of Porto, Porto, Portugal
[2]Centro de Química, Universidade de Tras-os-Montes e Alto Douro, Vila Real, Portugal
[3]Institute of Experimental Physics, SAS, Watsonova 47, 040 01 Kosice, Slovakia

**Abstract**

This work addresses the effect of substituting $Mn^{3+}$ by $Fe^{3+}$ at the octahedral site of $TbMnO_3$ on the magnetic phase sequence, ferroelectric and magnetoelectric properties, keeping the $Fe^{3+}$ concentration below 5%. The temperature dependence of the specific heat, dielectric permittivity and electric polarization was studied as a function of $Fe^{3+}$ concentration and applied magnetic field. From the experimental results a strong decrease of the electric polarization with increasing $Fe^{3+}$ substitution is observed, vanishing above a concentration of 4%. However, within this range, a significant increase of the magnetic sensitivity of the electric polarization is obtained by increasing $Fe^{3+}$ concentration. For $Fe^{3+}$ concentration above 4%, a non-polar phase emerges, whose spin structure prevent ferroelectricity according to the Dzyalowshinskii-Moriya model. The experimental results here reported reveal the crucial effect of B-site substitution on the magnetic phase sequence, as well as, on the polar and magnetoelectric properties, evidencing the important role played by the $e_g$ electrons on the stabilization of the magnetic structures, which are suitable for the emergence of electric polarization.

## 1. Introduction

The demand for new materials with higher performance and versatility has been focused on those exhibiting coupling between elementary excitations.[1,2] The transition metal oxides, with perovskite structure, have been largely studied because these materials exhibit strong coupling between orbital, electronic, spin and lattice degrees of freedom.[3] Among them, rare-earth manganites ($R$MnO$_3$, $R$ = trivalent rare-earth ion) offer a large variety of possibilities of tuning their structural and physical properties, namely by a deviation from stoichiometry.[4,5] In particular, the control of the magnetic structures and colossal magnetoresistance by atomic substitution has deserved much attention in the last decades from both theoretical and experimental point of views.[6–9] Among the possibilities of doping in perovskite transition metal oxides, the atomic substitution at the dodecahedral sites in stoichiometric compounds has been extensively studied.[5,10–13] However, only recently, interest has turned on to the effect of the substitution at the octahedral sites, where the experimental studies have been mainly focused on the effect on the magnetic behavior, and few on the multiferroic properties.[14–21] The effect of substitution at the octahedral sites with non-active Jahn-Teller ions on the ferroelectricity and magnetoelectric coupling still deserves attention, as the interpretation of the experimental results is controversial. This is the case of $TbMn_{1-x}M_xO_3$, with $M$ a trivalent transition metal ion: $Cr^{3+}$, $Fe^{3+}$ and $Co^{3+}$.[15–19] The lowest energy electronic states of these ions, under the influence of an octahedral crystal field and in the high spin configuration, are described by $t_{2g}^3e_g^0$, $t_{2g}^3e_g^2$ and $t_{2g}^4e_g^2$/(or $t_{2g}^6e_g^0$, for $Co^{3+}$ in the low spin state, also found in orthocobaltites), respectively, while for the $Mn^{3+}$ ion is $t_{2g}^3e_g^1$. Among the aforementioned ions, only $Mn^{3+}$ exhibits the Jahn-Teller effect. It was shown that the substitution of $Mn^{3+}$ by $Cr^{3+}$ for $x$ up to 0.33 induces ferromagnetic-like Mn-Cr interactions that are enhanced by an applied magnetic field and promote a ferromagnetic character of the solid solution with increasing $x$ from 0.5 up to 1.[19] Similar findings were reported for the $TbMn_{1-x}Co_xO_3$ solid solution ($0.1 \leq x \leq 0.9$), where the enhancement of ferromagnetic correlations were associated with the existence of magnetically inhomogeneous clusters arising from competitive magnetic interactions.[18] Very recently, the effect of the $Mn^{3+}$ substitution by $Cr^{3+}$ was experimentally ascertained to gradually suppress the ferroelectric phase, stabilized by the $Mn^{3+}$ cycloidal spin ordering.[17] Based on the Dong *et al* model,[22] which highlights the key role played by the $e_g$ electrons in forming the cycloidal spin order, the Gou *et al* have interpreted the loss of the ferroelectric phase by the reduction of the number of $e_g$ electrons in this system.[17] The suppression of ferroelectricity was also reported in $TbMn_{1-x}Ru_xO_3$ for $x \geq 0.1$, which has been assigned to the disruption of the $Mn^{3+}$ spiral spin order, reduction of the Mn-O-Mn angle and quantum fluctuations.[21] Unfortunately, to the best of our knowledge, no similar works were published on other ionic substitution at the octahedral site of $TbMnO_3$.

The phase sequence of $TbMnO_3$ is well reported in the literature, and can be summarized as follows.[23,24] At $T_N$ = 41 K, $TbMnO_3$ undergoes a magnetic phase transition into a incommensurate antiferromagnetic phase, with a longitudinal spin density wave propagating along the *a*-axis, in a P*bnm* symmetry setting.[25,26] Below $T_{lock}$ = 27 K, a cycloidal spin order in the *bc*-plane becomes stable, which breaks the space inversion symmetry and thus induces a spontaneous electric polarization along the *c*-axis, according to the Dzyaloshinskii-Moriya mechanism.[25,26] On further cooling, $Tb^{3+}$ spins order at $T_1$ = 7 K in the effective magnetic field originated from the Mn sublattice.[26]

The inclusion of a magnetic non-active Jahn-Teller $Fe^{3+}$ cation in the octahedral sites of $TbMnO_3$, even in small concentrations, is a very effective route to reach a substantial change of its magnetic properties.[27] Due to the fact that both $Mn^{3+}$ and $Fe^{3+}$ ions have the same radius, a fully solubility is obtained.[14–16] As the magnetic $Mn^{3+}$ ion is being substituted by another magnetic ion, the magnetic phase diagram is expected to be dependent on $Fe^{3+}$ substitution.[15,28] In fact, for $Fe^{3+}$ concentration up to around 50%, a weak-ferromagnetic phase emerges, wherein $T_N$ increases from 41 K to 300 K.[14,15,28] For higher concentrations, a continuous increase of the $T_N$ line is observed up to 650 K, whereas at lower temperatures a spin order reorientation occurs below a transition temperature that decreases from room temperature down to 10 K.[14,15,28] Very recently, powder neutron diffraction studies in Fe-substituted $TbMnO_3$ has shown that the Mn long-range incommensurate spin ordering is destroyed by the introduction of $Fe^{3+}$ for concentrations of 10% to 25%, giving rise to a ferromagnetic component along the *b*-axis resulting in a highly frustrated spin state.[29,30] It is worth noting that none of the aforementioned works include a detailed study of the low concentration range of $Fe^{3+}$, below 10%, namely the effect of the $Fe^{3+}$ dopant on the magnetic phase sequence, ferroelectric and magnetoelectric properties. This range of $Fe^{3+}$ concentration is from our standpoint of view a crucial feature, regarding the polar properties and magnetoelectric coupling.

As it was referred to above, the explanation of the effect of the B-site substituent on the magnetic and polar properties of $TbMnO_3$ is still under debate. This work aims at unraveling the effect of $Fe^{3+}$ substitution, up to *x* = 0.05, in the magnetic, ferroelectric and magnetoelectric properties of the $TbMn_{1-x}Fe_xO_3$ system, both as a function of temperature and applied magnetic field, up to 9 T. The partial substitution of $Mn^{3+}$ by $Fe^{3+}$, exhibiting different electronic configurations of the *d*-orbitals in the high spin state as aforementioned, along with the available data concerning the effect of the substituent on the electric polarization, will be an important challenge, since it will allow a deeper understanding of the role of the $e_g$ electrons on stabilizing the spiral spin order of TbMnO$_3$. Particular attention will be paid to the stability range of the magnetically induced ferroelectric phase and its sensitivity to an applied magnetic field. The results are discussed in the scope of available theoretical framework, with special focus on the role of the $e_g$ electrons and their non-degenerated ground-states.

## 2. Experimental details

A series of high-quality TbMn$_{1-x}$Fe$_x$O$_3$ ($x$ = 0.0, 0.01, 0.025, 0.04 and 0.05) ceramics were made by urea sol-gel combustion method,[31] whose chemical, morphological and microstructural properties were fully characterized. Specific heat was measured using a relaxation technique incorporated in a PPMS and the complex electric permittivity with an impedance analyzer. The pyroelectric currents were obtained with a standard short-circuit method, integrated in a SQUID magnetometer insert for the applied magnetic field. For more details see Ref. [12].

## 3. Experimental results and discussion

As the ($x$, T) phase diagram of the solid solution is unknown in the 0.0 < $x$ < 0.1, the $x$-dependence of the critical temperatures of the TbMn$_{1-x}$Fe$_x$O$_3$ system, at zero magnetic field, was obtained from the temperature dependence of specific heat divided by temperature (C/T) and electric permittivity, which are shown in Figures 1(a) and (b), respectively, for the representative cases of $x$ = 0.0, 0.025 and 0.05. The three anomalies observed in both C/T and electric permittivity curves of TbMnO$_3$ are associated with the following transitions to: (i) the collinear sinusoidal incommensurate modulated antiferromagnetic (IC-AFM) phase at $T_N$ = 41 K; (ii) the cycloidal commensurate modulated antiferromagnetic (C-AFM) phase at $T_{lock}$= 25 K (marked by *), which is also ferroelectric; (iii) the modulated quasi-long range antiferromagnetic ordering of the Tb$^{3+}$ ions at $T_1$ = 7 K. The experimental values obtained for the critical temperatures are in good agreement with those reported in the literature.[24,32] Based on the phase sequence observed in TbMnO$_3$ and the existence of the three anomalies in the temperature dependence of C/T and $\varepsilon'$, we assign the same phase sequence as TbMnO$_3$ for the compositions with $x \leq 0.04$. The anomaly at T$_1$ barely shifts with Fe-substitution. This result is similar to that observed in TbMn$_{1-x}$Cr$_x$O$_3$ ($x \leq 0.06$) ceramics, and corroborates that the *f-d* exchange interaction between the Mn 3*d* spins and the Tb 4*f* moments in multiferroic TbMnO$_3$ almost involves only the $t_{2g}$ electrons.[24] For increasing $x$ up to 0.025, both T$_N$ and T$_{lock}$ shift to lower temperatures, and the three anomalies are detected in both C/T(T) and $\varepsilon'$(T). For $x$ = 0.05, we clearly only observe two anomalies in the C/T(T) curve, at 35 K and at 6 K, respectively. The decrease of both T$_N$ and T$_{lock}$ with increasing $x$ up to 0.04 evidences for the weakening of the antiferromagnetic interactions and thus, the energetic balance makes these phases stable at lower temperatures. However, for Fe$^{3+}$ concentrations above 4%, the anomaly associated with T$_{lock}$ is no longer perceived in electric permittivity and specific heat curves. We interpret this result as the suppression of the cycloidal spin ordered phase for $x$ > 0.04, which was evidenced by powder neutron diffraction measurements for $x$ = 0.1[29] The $x$-dependence of the critical temperatures, along with the identification of the magnetic phases, are summarized in the ($x$, T) phase diagram shown in Figure 1(c).

In order to unravel the effect of the Mn$^{3+}$ substitution by Fe$^{3+}$ on the polar nature of the magnetic phase observed below T$_{lock}$, the temperature dependence of the pyroelectric current was studied for different $x$-values. Figure 2(a) shows the temperature dependence of the pyroelectric current density for several values of $x$, and Figure 2(b) refers to the temperature dependence of the electric polarization, calculated from time integration of pyroelectric current density. As Fe$^{3+}$ concentration increases, it is observed that, besides the decrease of T$_{lock}$, the polarization at 10 K strongly decreases.

The decrease of the electric polarization is associated with the gradual fading out of cycloidal spin ordering, whose disappearance prevents the occurrence of ferroelectricity in agreement with the Dzyaloshinskii-Moriya mechanism. This feature is similar to the TbMn$_{1-x}$Cr$_x$O$_3$ system and reveals the strong effect on the superexchange interactions due to the different $e_g$ electron configurations of the B-site cation on the stabilization of the cycloidal modulated structure, as it will be explain in the following.

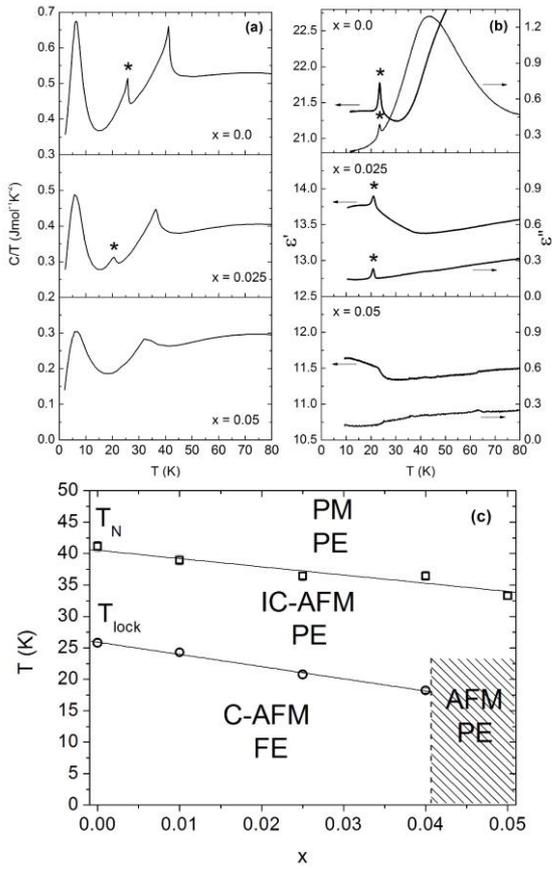
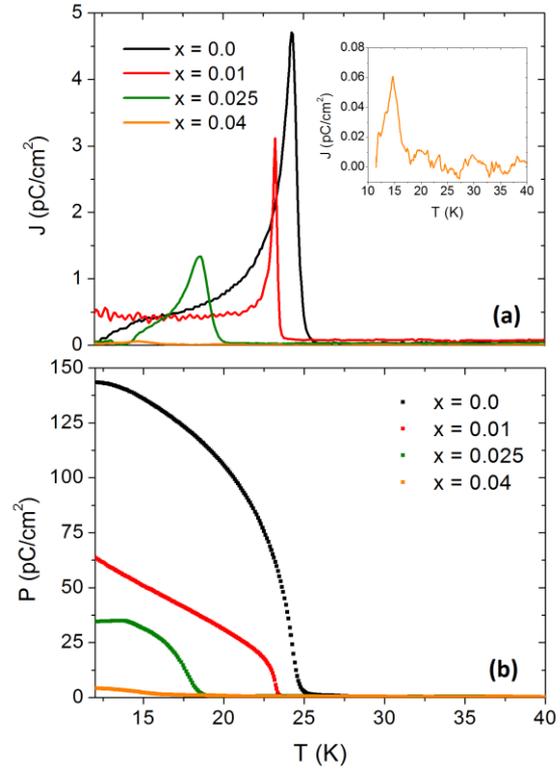

Figure 1. (a) Specific heat divided by temperature and (b) real and imaginary part of dielectric permittivity, measured in heating runs at 85.7 kHz, for $x$ = 0.0, 0.025 and 0.05. Star (*) marks $T_{lock}$. (c) Magnetic ($x$, $T$) phase diagram of TbMn$_{1-x}$Fe$_x$O$_3$, where IC-AFM stands for sinusoidal incommensurate modulated antiferromagnetic and C-AFM for cycloidal commensurate antiferromagnetic. The grey area is a guess for the AFM-paraelectric phase.

Figure 2. (a) Pyroelectric current density and (b) respective electric polarization curves as a function of $x$. The inset of figure (a) shows a detailed view of the temperature dependence of the pyroelectric current measured from the sample with $x$ = 0.04.

The most suitable theoretical model to describe the phase diagram of the rare-earth manganites, proposed by M. Mochizuki and N. Furukawa, is based on the competition between nearest neighbor Mn$^{3+}$ ferromagnetic exchanges ($J_{ab}$) and next nearest neighbor antiferromagnetic exchanges ($J_2$).[8,9] The $J_{ab}$ superexchange has a contribution arising from the ferromagnetic interactions of the staggered $e_g$ electrons in the $ab$-plane, while the $J_2$ has the contribution of the antiferromagnetic arrangement of the $e_g$ electrons.[8] The $t_{2g}$ electrons give always an antiferromagnetic contribution to both superexchange interactions, which we will take as independent for low $x$ values.[8,9]

In rare-earth manganites, the two $J_{ab}$ and $J_2$ exchanges are comparable in strength, and the next nearest neighbor $J_2$ exchange plays important role in determining the magnetic structure.[8] The cycloidal spin ordering in TbMnO$_3$ is stabilized through the frustration of the competing ferromagnetic and antiferromagnetic interactions.[8] The main factor driving the magnetic modifications should be that the $Fe^{3+}$ introduction locally alters both the number of the $e_g$ electrons and the Jahn-Teller distortion, in such a way that the local nearest neighbor ferromagnetic interaction of the $e_g$ electrons should increase as the next neared neighbor antiferromagnetic one decreases, and thus, enhancing $J_{ab}$.[8,26,29] Similar ferromagnetic distortion of the spiral spin order was also reported arising from Mn-Cr interactions in the TbMn$_{1-x}$Cr$_x$O$_3$ system.[19] For $x > 0.04$, the ratio $J_2/J_{ab}$ should become small enough that the cycloidal spin order is no longer stable.

Now, we explore how the ($x$, T) phase diagram and the magnetoelectric coupling strength evolve under applied magnetic field. To reach these objectives, the temperature dependence of specific heat and pyroelectric current were studied under applied magnetic fields, for different $x$ values. Figures 3(a) to (c) show representative examples of the temperature dependence of C/T for $x$ = 0.0, 0.01, and 0.05, under different magnetic fields up to 9 T. It is worth mentioning that, for every studied composition, a large anomaly of C/T is observed at $T_1$ = 7 K for zero applied magnetic field, which can sign the $Tb^{3+}$ spin ordering occurring below to that temperature. For increasing magnetic field the anomaly becomes a hump, and likely disappears for high enough field strength. The nature of this behavior is still a matter under current debate, and being out of scope of the aim of this work, will not be further addressed. The anomalies at $T_{lock}$, and $T_N$ are much more stable with respect to the applied magnetic field and in generally their response on magnetic field is not monotonous. From the $x$, magnetic field and temperature dependences of the anomalies in the C/T curves a ($x$, $\mu_0$H, T) phase diagram could be traced, which is presented in Figure 3(d).

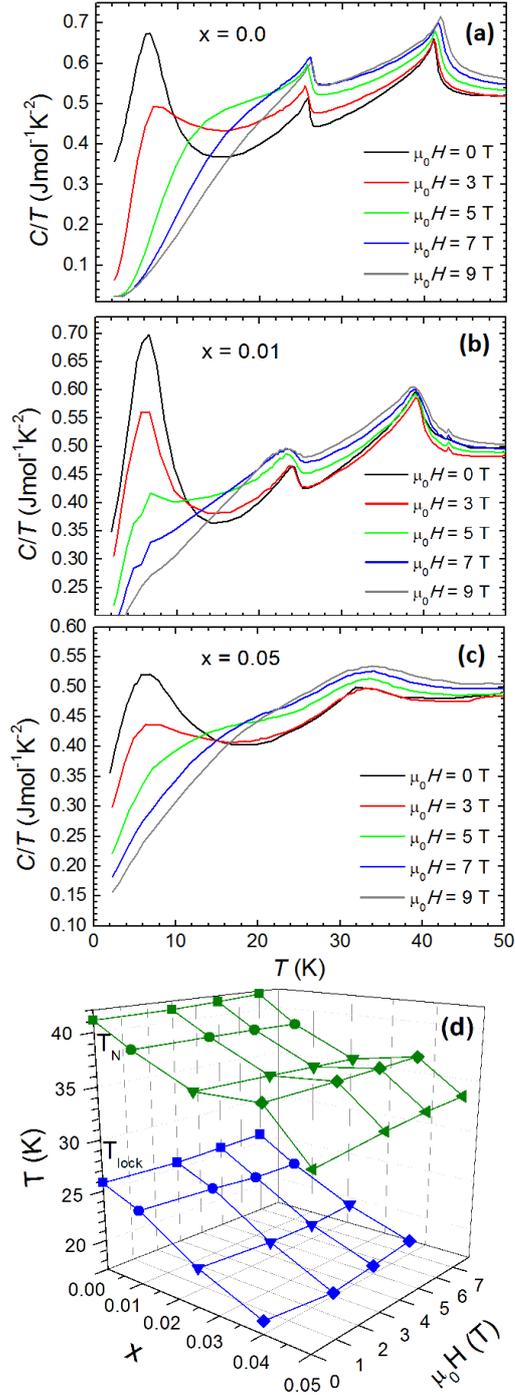

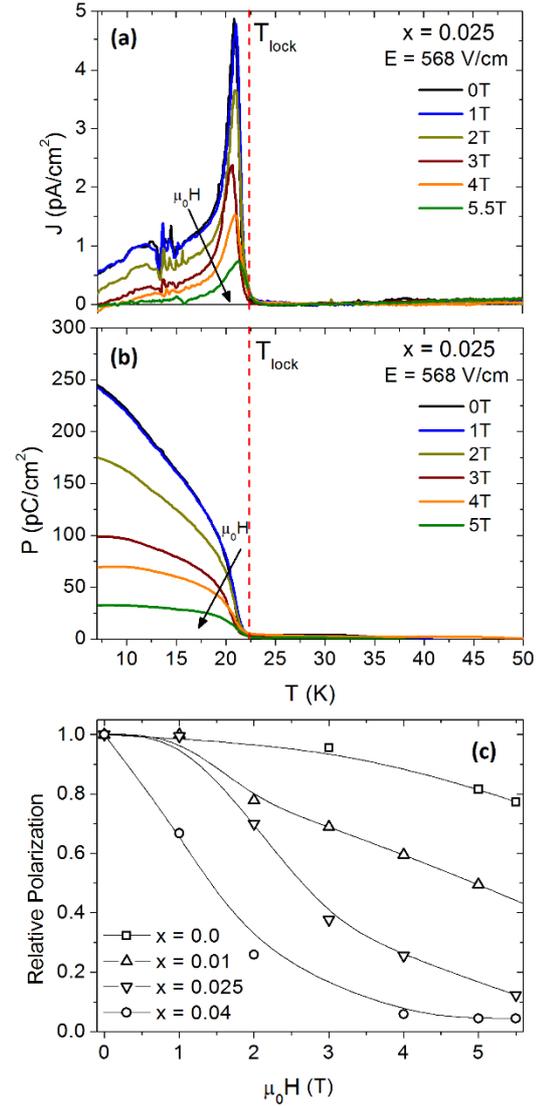

Figure 3. C/T temperature dependence of TbMn$_{1-x}$Fe$_x$O$_3$, for the cases of (a) x = 0.0, (b) 0.01 and (c) 0.05. (d) T$_N$ and T$_{lock}$ phase surfaces dependencies with Fe-concentration and applied magnetic field.

Figure 4. (a) Example of the pyroelectric current density and (b) corresponding eletric polarization curves as a function of temperature under several magnetic fields for x = 0.025. Dashed lines mark T$_{lock}$. (c) Relative electric polarization, P$_{rel}$, at 5 K, with poling electric field of around 500 V/cm, as function of applied magnetic field for x = 0.0 to 0.04. Lines are guides for the eyes.

Figures 4(a) and (b) show, for x = 0.025, pyroelectric current density and corresponding electric polarization as a function of temperature, for magnetic fields up to 5.5 T, respectively. Moreover, Figure 4(c) shows the magnetic field dependence of the relative electric polarization ($P_{rel}$) for x = 0.0, 0.01, 0.025 and 0.04, defined as: $P_{rel} = \frac{P(\mu_0 H)}{P(0)}$, where P(0) is the polarization measured at 5 K without magnetic field, and P($\mu_0$H) is the polarization measured at 5 K under magnetic field. In the case of TbMnO$_3$, no apparent change in the relative electric polarization curve is observed for applied magnetic fields below to 1 T. Furthermore, it decreases only to 85% of its initial value up to 5 T. The magnetic field dependence of the relative electric polarization is in good agreement with the magnetic dependence of the electric polarization in TbMnO$_3$ obtaining by "summing up" the *a* and *c*-axis polarizations as a function of the magnetic field,[23] wherein the former increases and the latter decreases with increasing applied magnetic field.

As x increases, the sensitivity of the relative electric polarization to the applied magnetic field drastically increases, as it can be observed in Figure 4(c). For the case of x = 0.04, a third of the relative electric polarization is already removed at a magnetic field of 1 T, and more than 95% of its value can be suppressed by applying a magnetic field above 4 T. Figure 4(c) clearly evidences that the inclusion of the Fe$^{3+}$ ions has strong effect on sensibility of the electric polarization on the applied magnetic field. This means that the magnetoelectric coupling is strengthened by the Fe-substitution.

The increase of sensitivity of the electric polarization to an applied magnetic field with increasing *x*, is understood by the magnetic field enhancement of the Fe$^{3+}$-induced ferromagnetic contribution of the $e_g$ electrons to the $J_{ab}$ exchange, which also explains the similar magnetic field dependence of the electric polarization found in TbMn$_{1-x}$Cr$_x$O$_3$.[17] In both systems, the $e_g$ electrons of the substituent magnetic ions increases the ferromagnetic $J_{ab}$ against the antiferromagnetic $J_2$ one, evidencing their important role underlying the magnetoelectric coupling in rare-earth manganites.

## 4. Conclusions

In summary, this work focused on the effect of the $e_g$ electrons introduced by the substitution of Jahn-Teller active magnetic Mn$^{3+}$ by the non-active Jahn-Teller active magnetic Fe$^{3+}$ ions, on the magnetic, ferroelectric and magnetoelectric properties of TbMn$_{1-}$

$_x$Fe$_x$O$_3$, for $x$ up to 0.05. The ($x$, $\mu_o$H, T) phase diagram could be traced from the experimental results obtained by specific heat, dielectric permittivity and polar measurements. One of the most significant outcomes is the strong decrease of the electric polarization with increasing Fe$^{3+}$ concentration, disappearing for $x > 0.04$. We propose that the transition to this phase can be understood on the basis of the increase of the nearest neighbor ferromagnetic J$_{ab}$ exchanges with the concomitant decrease of the next nearest neighbor antiferromagnetic J$_2$ exchanges, which is induced by the local interaction of the e$_g$ electrons of the Fe$^{3+}$ substituent with the Mn$^{3+}$ ones. This interpretation is also in agreement with the decrease of T$_N$ and T$_{lock}$ with $x$, which indicates that the corresponding magnetic orderings are becoming less stable. Another important outcome is the large change of the low field magnetoelectric response through Fe$^{3+}$ substitution, associated with the increase of the electric field sensitivity to the applied magnetic field. It is experimentally shown that for both Fe$^{3+}$ and Cr$^{3+}$ substituents, a monotonous decrease of the relative electric polarization, though the rate of decrease is apparently larger for Fe$^{3+}$ substituted system. All these findings point to a specific role played by the $e_g$ electrons on the nearest neighbor ferromagnetic coupling. In this framework, Mn-Fe bonds are expected to show larger values of $J_2$ relatively to those of Mn-Cr.

## Acknowledgments

The authors would like to acknowledge the support of the projects Norte-070124-FEDER-000070, UID/NAN/50024/2013, PTDC/FIS-NAN/0533/2012, ERDF EU contract no. ITMS26220120047 and VEGA2/0132/16, and R. Vilarinho to the grant PD/BI/106014/2015 by FCT.